\documentclass[a4paper,11pt]{article}
\pdfoutput=1 

\usepackage{jcappub} 
\usepackage{indentfirst}
\usepackage{appendix}
\usepackage{multirow}

\title{Relieving the $H_0$ tension with a new interacting dark energy model
}

\author[a]{Li-Yang Gao,}
\author[a]{Ze-Wei Zhao,}
\author[b,c]{She-Sheng Xue,}
\author[a,1]{Xin Zhang\note{Corresponding author.}}

\affiliation[a]{Department of Physics, College of Sciences, \& MOE Key Laboratory of Data Analytics and Optimization
for Smart Industry, Northeastern University, Shenyang 110819, China}
\affiliation[b]{ICRANet, Piazzale della Repubblica, 10-65122, Pescara}
\affiliation[c]{Physics Department, Sapienza University of Rome, P.le A. Moro 5, 00185, Rome, Italy}

\emailAdd{1910025@stu.neu.edu.cn, 1810024@stu.neu.edu.cn, shesheng.xue@gmail.com, zhangxin@mail.neu.edu.cn}

\abstract{We investigate an extended cosmological model motivated by the asymptotic safety of gravitational field theory, in which the matter and radiation densities and the cosmological constant receive a correction parametrized by the parameters $\delta_G$ and $\delta_\Lambda$, leading to that both the evolutions of the matter and radiation densities and the cosmological constant slightly deviate from the standard forms. Here we explain this model as a scenario of vacuum energy interacting with matter and radiation. We consider two cases of the model: {(i) ${\tilde\Lambda}$CDM with one additional free parameter $\delta_G$, with $\delta_{\rm G}$ and $\delta_\Lambda$ related by a low-redshift limit relation and (ii) e${\tilde\Lambda}$CDM with two additional free parameters $\delta_G$ and $\delta_\Lambda$ that are independent of each other.} We use two data combinations, CMB+BAO+SN (CBS) and CMB+BAO+SN+$H_0$ (CBSH), to constrain the models. We find that, in the case of using the CBS data, neither ${\tilde\Lambda}$CDM nor e${\tilde\Lambda}$CDM can effectively alleviate the $H_0$ tension. However, it is found that using the CBSH data the $H_0$ tension can be greatly relieved by the
models. In particular, in the case of e${\tilde\Lambda}$CDM, the $H_0$ tension can be resolved to 0.71$\sigma$. We conclude that as an interacting dark energy model, ${\tilde\Lambda}$CDM is much better than $\Lambda(t)$CDM in the sense of both relieving the $H_0$ tension and fitting to the current observational data.}

\begin{document}
\maketitle
\flushbottom

\section{Introduction}
\label{sec1}

The Hubble constant $H_0$ is the first cosmological parameter, which was introduced by Edwin Hubble to describe the current expansion of the universe, and it has been measured for about one century. Precisely measuring the value of the Hubble constant is extremely important for cosmology because it determines the absolute scale of the universe.
But with the development of precision cosmology, cosmologists now face an increasingly puzzling problem, i.e., the discrepancy between the value of $H_0$ inferred from the early universe using the cosmic microwave background (CMB) data observed by the {\it Planck} satellite assuming a base $\Lambda$CDM cosmology \cite{Aghanim:2018eyx} and the one directly measured by using the Cepheid-supernovae distance ladder \cite{Riess:2019cxk}.
Based on the CMB measurements from {\it Planck} TT,TE,EE+lowE+lensing \cite{Aghanim:2018eyx} and baryon acoustic oscillation (BAO) measurements from galaxy redshift surveys \cite{Beutler:2011hx,Ross:2014qpa,Alam:2016hwk}, it is found that in the base $\Lambda$CDM model we have $ H_0 = (67.36\pm 0.54)~\rm{km~s^{-1}~Mpc^{-1}}$ \cite{Aghanim:2018eyx}.
On the other hand, the direct measurement of the Hubble constant from the {\it Hubble Space Telescope} using the distance ladder method gives the result of $ H_0 = (74.03\pm 1.42)~\rm{km~s^{-1}~Mpc^{-1}}$, which shows a $4.4\sigma$ tension in statistical significance with the early-universe result from the {\it Planck} CMB measurement (for some reviews on this tension, see Refs. \cite{Verde:2019ivm,DiValentino:2021izs,DiValentino:2017gzb,DiValentino:2020zio,Freedman:2017yms,Riess:2020sih}).
The reasons for this tension are usually ascribed to systematic errors or new physics.

To solve this problem, a number of articles have attempted to address the systematic errors in these two methods \cite{Spergel:2013rxa,Addison:2015wyg,Aghanim:2016sns,Efstathiou:2013via,Cardona:2016ems,Zhang:2017aqn,Follin:2017ljs}, but no reliable evidence is found and the tension actually still exists.
Therefore, it is of great importance to measure the Hubble constant in other independent ways.
In fact, besides the Cepheid-supernova distance ladder, there are also two distance ladders, i.e., the ones using Mira variables \cite{Huang:2019yhh} or red giants \cite{Yuan:2019npk} instead of Cepheids to calibrate type Ia supernovae (SNIa).
Other late-universe measurement methods also include the observations of strong lensing time delays \cite{Wong:2019kwg}, water masers \cite{Pesce:2020xfe}, surface brightness
fluctuations \cite{Jensen:1998bi}, gravitational waves from neutron star mergers \cite{Abbott:2017xzu}, different ages of galaxies as cosmic clocks \cite{Jimenez:2001gg,Moresco:2016mzx}, baryonic Tully-Fisher relation \cite{Schombert:2020pxm}, and so forth.
All these observations show that the late-universe estimations of $H_0$ disagree with the prediction from the {\it Planck} CMB observation in conjunction with the base $\Lambda$CDM cosmology at the 4--6$\sigma$ level.

On the other hand, there have been lots of theoretical ideas \cite{Li:2013dha,Camarena:2018nbr,Salvatelli:2013wra,Costa:2013sva,DiValentino:2017iww,Yang:2017ccc,DiValentino:2021zxy,Yang:2021hxg,Zhao:2017cud,Qing-Guo:2016ykt,Martinelli:2019krf,Alestas:2020mvb,DiValentino:2020vnx,Efstathiou:2020wxn,Yang:2021flj} to address the Hubble tension by extending the standard model of cosmology.
For example, in the aspect of the late universe, one may consider dynamical dark energy instead of the cosmological constant, or the interaction between dark energy and dark matter, and in the aspect of the early universe, one may consider the extra relativistic degrees of freedom, early dark energy, or the self-interaction among neutrinos.
A comprehensive analysis of many typical extended cosmological models \cite{Guo:2018ans} shows that among these extended models actually no one can truly resolve the Hubble
tension.

In this paper, we wish to investigate a new extension to the standard $\Lambda$CDM model, which is motivated by the asymptotic safety of gravitational field theory \cite{Xue:2014kna}, from the perspective of how to relieve the $H_0$ tension.
As the universe expands and the energy (time) scale varies, the gravitational coupling parameter \textit{G} and the cosmological constant $\Lambda$ will vary following scaling laws and approach to the present values $G_0$ and $\Lambda_0$.
This implies that in the normal Friedmann equations of $\Lambda$CDM the matter (radiation) term $\Omega_{\rm m,r}$ and the cosmological constant term $\Omega_{\Lambda}$ could receive an additional scaling factor $(1+z)^{\delta}$ with $\delta\ll 1$.
To constrain the model parameter $\delta$ and address the $H_0$ tension issue, we adopt the combination of the latest cosmological datasets $\rm CMB+BAO+SN$ with or without the $H_0$ prior from the local measurement, compared to the $\Lambda$CDM model and some other typical cosmological models.
In our analysis, we fit all the models to the same datasets and examine the $H_0$ tension by taking $\Lambda$CDM as a benchmark model.

The structure of this paper is arranged as follows.
In Section \ref{sec2}, we present the description of the new extended cosmological model.
Section \ref{sec3} briefly describes the data and methods used in this work.
The results and related analysis are presented in Section \ref{sec4}.
We test the robustness of results in Section \ref{sec5}.
Conclusion is given in Section \ref{sec6}.

\section{Motivation and cosmological models}\label{sec2}

The $\Lambda$CDM model has usually been viewed as a standard model of cosmology at the
present.
In the $\Lambda$CDM model, the expansion history of the universe, described by the
Hubble expansion rate, is given by the Friedmann equation,
\begin{eqnarray}
H^2= \frac{8\pi G}{3}(\rho_{\rm m}+\rho_{\rm r}+\rho_{\rm \Lambda}),
\label{hubble}
 \end{eqnarray}
where $H$ is the Hubble parameter, $G$ is the gravitational constant, and the densities
of matter and radiation evolve with redshift as $\rho_{\rm m,r} = \rho^0_{\rm m,r}(1+z)^{3(1+w_{\rm m,r})}$, with their equations of state $w_{\rm m}= 0$ for non-relativistic particles and $w_{\rm r}= 1/3$ for relativistic particles.
The cosmological constant $\Lambda$ describes the vacuum energy density, which serves as dark
energy in this model.
The vacuum energy density is given by $\rho_{_\Lambda} = \rho^0_{_\Lambda}\equiv \Lambda/(8\pi G_0)$, which has a negative pressure with the equation of state $p_{_\Lambda} = w_{_\Lambda} \rho_{_\Lambda}$, with $w_{_\Lambda} = -1$. Note that here in fact we use $\Lambda$ to denote the ``effective'' cosmological constant $\Lambda\simeq 4.2\times 10^{-66}~{\rm eV}^2=2.8\times 10^{-122}~m_{\rm Pl}^2$ with $m_{\rm Pl}$ the Planck mass. Actually, the puzzling problem of why the original vacuum energy density could precisely cancel with the ``bare'' cosmological constant leading to such a small value of $\Lambda$ is still an open question, also known as the cosmological constant problem, which is usually viewed to be closely relevant to quantum gravity, and we will not deeply discuss this issue in this paper.

Here we present a new extended cosmological model.
The principle of the new model discussed in this work is the same as in Ref.~\cite{Xue:2014kna}, and we assume that the gravitational constant varies with redshift.
As a consequence, the cosmological constant $\Lambda$ will also change with the redshift because of this assumption.
In this paper, the quantities with
subscript or superscript ``$0$'' stand for their present values ($z=0$), i.e., $G_0$ and $\Lambda_0$ are the present values of gravitational constant and cosmological constant, respectively, while $\rho^0_{\rm m}$, $\rho^0_{\rm r}$, and $\rho^0_{_\Lambda}$ are the densities of matter, radiation, and dark energy at the present, respectively.

As one of the fundamental theories for interactions in nature, the classical Einstein theory of gravity, which plays an essential role in the  standard model of modern cosmology ($\Lambda$CDM), should be  realized in the scaling-invariant domain of a fixed point of its quantum field theory. It was suggested by Weinberg \cite{Weinberg:2009wa} that the quantum field theory of gravity regularized with an ultraviolet (UV) cutoff might have a non-trivial UV-stable fixed point and asymptotic safety, namely the renormalization group (RG) flows are attracted into the UV-stable fixed point with a finite number of physically renormalizable operators for the gravitational field.
Ref.~\cite{Xue:2014kna} studied the asymptotic safety of the quantum field theory of gravity, 
namely the gravitational ``constant'' $G$ and the cosmological ``constant'' $\Lambda$ are time varying, approaching to the point $(G_0,\Lambda_0)$ where two relevant operators of Ricci scalar term \textit{R} and cosmological term $\Lambda$ of classical Einstein gravity are realized.
This implies the ``scaling laws'' (ansatz) $G/G_0=(1+z)^{-\delta_{\rm G}}$ and $\Lambda/\Lambda_0=(1+z)^{\delta_\Lambda}$, where the two ``critical exponents'' (parameters)
$\delta_{\rm G} \ll 1$ and $\delta_\Lambda \ll 1$ are related.
This motivates us to extend the $\Lambda$CDM model by assuming
\begin{eqnarray}
(G/G_0)\rho_{\rm m,r} &=  & \rho^0_{\rm m,r}(1+z)^{3(1+w_{\rm m,r})-\delta_G},
\label{moa}\\
 (G/G_0)\rho_{_\Lambda} &= & \rho^0_{_\Lambda}(1+z)^{+\delta_\Lambda},\label{lambda}
 \end{eqnarray}
where $w_{\rm m}\approx 0$, $w_{\rm r } \approx 1/3$, and $\rho_{_\Lambda}\equiv \Lambda/(8\pi G)$ is time varying, but $w_{_\Lambda}= -1$ still holds. The parameter $\delta_G$ is the same for the matter $\rho_{\rm m}$ and radiation $\rho_{\rm r}$ terms,
assuming the deviation is only due to time-varying $G$.
Two Friedmann equations are extended to
\begin{eqnarray}
E^2(z) &=&\Omega_{\rm m}(1+z)^{(3-\delta_{\rm G})} + \Omega_{\rm r}(1+z)^{(4-\delta_{\rm G})}+\Omega_{_\Lambda}(1+z)^{\delta_\Lambda},
\label{final}\\
(1+z)\frac{d}{dz}E^2(z)
&=&3\Omega_{\rm m}(1+z)^{(3-\delta_{\rm G})}+4\Omega_{\rm r}(1+z)^{(4-\delta_{\rm G})},
\label{fcgeqi20}
\end{eqnarray}
where $E(z)\equiv H(z)/H_0$. Here, $\Omega_{\rm m}$, $\Omega_{\rm r}$, and $\Omega_{_\Lambda}=1-\Omega_{\rm m}-\Omega_{\rm r}$ are the present-day fractional energy densities of matter, radiation, and dark energy, respectively.
Eq.~(\ref{fcgeqi20}) comes from the generalized energy conservation law \cite{Xue:2014kna} for varying gravitational and cosmological ``constants" interacting with matter and radiation. It reduces to the matter conservation in usual Friedman equations for constants $\Lambda$ and $G$. Substituting Eq.~(\ref{final}) to Eq.~(\ref{fcgeqi20}), we find the relation of the parameters $\delta_{\rm G}$ and $\delta_\Lambda$,
\begin{eqnarray}
\delta_\Lambda &\approx & \delta_{\rm G}\left(\frac{\Omega_{\rm m}+\Omega_{\rm r}}{\Omega_{_\Lambda}}\right)\approx 0.47 ~\delta_{\rm G},
 \label{deltal}
\end{eqnarray}
in the low redshift ($z\rightarrow 0$) limit. Nonzero $\delta_{\rm G,\Lambda}$ show that dark energy and matter interact and can be converted from one to another. They obey the total energy conservation (\ref{fcgeqi20}). The relations of small parameters $\delta_{\rm G,\Lambda}$ to other interacting models of dark energy and matter can be found in Eqs.~(10)--(15) of
Ref.~\cite{Begue:2017lcw}.

Notwithstanding the absence of the detailed and explicit interpretation of such a modelling $E(z)$, we are in the position of providing some insights into possible physics.
The parameters $\delta_{\rm G}$ and $\delta_\Lambda$ effectively represent the possible physical effects or combinations of these effects in addition to those of the $\Lambda$CDM model, such as: small time-varying gravitational constant $G$ and inhomogeneity of matter distribution in different redshift $z$; the transition from the radiation-dominated era to the matter-dominated era, and {\it vice versa}, depending on species of normal particles or dark matter particles;
and massive particle production and annihilation due to the interaction between dark energy (vacuum energy) and other cosmological components \cite{Xue:2020tpf,Xue:2019otx}. $\delta_{\rm G}> 0$ or $\delta_{\rm G} < 0$ implies that the decrease of $\rho_{\rm m,r}$ is slower or faster than that of $\Lambda$CDM.
Actually, we can treat the model as a kind of interacting dark energy (vacuum energy) model, and
thus the effects of $\delta_{\rm G}\not= 0$ and $\delta_\Lambda \not= 0$ in the late universe are expected.
Here, we wish to emphasize the usage of the terminology of ``vacuum energy'' in the following of this work; actually we exactly refer to ``vacuum energy'' with the case of $w=-1$.

In general, the value of the parameter $\delta_G$ can be different for matter ($\rho_{\rm m}$) and radiation ($\rho_{\rm r}$) terms in $E^2(z)$ in Eq.~(\ref{final}), since dark energy should interact differently with matter and radiation. Therefore, we consider in this article two cases: (i) $\delta_{\rm G,\Lambda}$ related by the relation Eq. (\ref{deltal}) and (ii) $\delta_{\rm G,\Lambda}$ independent from each other.
Henceforth, for a short
notation and readers' convenience, the one-parameter extended
model  for the first case with the relation
(\ref{deltal}) is called the ``varying $\Lambda$''CDM, represented by the symbol ${{\tilde\Lambda}}$CDM. Whereas, because the second case has one more parameter than the ${{ \tilde\Lambda}}$CDM model, the two-parameter extension is called the extended ${{ \tilde\Lambda}}$CDM, also abbreviated as e${{ \tilde\Lambda}}$CDM.



In this article, we compare the ${\tilde\Lambda}$CDM model with other one-parameter extensions of the $\Lambda$CDM model, i.e., $w$CDM and $\Lambda(t)$CDM.  Besides, we compare the e${\tilde\Lambda}$CDM model with the Chevallier-Polarski-Linder (CPL) model, both are two-parameter extensions to $\Lambda$CDM.
These models used for comparison are summarized as follows:

1. the $w$CDM model \cite{Chevallier:2000qy,Linder:2002et}: The equation-of-state parameter $w$ is treated as a constant free parameter instead of $w= -1$. We adopt $E^2(z) = \Omega_{\rm m}(1+z)^{3} +\Omega_{\rm r}(1+z)^{4}+ \Omega_{_\Lambda}(1+z)^{3(1+w)}$.

2. the $\Lambda (t)$CDM model \cite{Guo:2017hea,Feng:2017usu,Guo:2018gyo}: The vacuum energy with $w_{_\Lambda}=-1$ serves as dark energy, and the interaction between dark energy (vacuum energy) and cold dark matter is described by the equations $\dot\rho_{\rm de} = Q$ and $\dot\rho_{\rm c} = -3H\rho_{\rm c} - Q$.
Here, the subscript ``de" is for dark energy and the subscript ``c" is for cold dark matter.
The interaction term $Q = -\beta H \rho_{c}$ determines characteristics of energy transfer between dark energy and dark matter, and $\beta$ is a dimensionless coupling parameter.

3. the CPL model \cite{Chevallier:2000qy,Linder:2002et}: We have $w(a)=w_0+w_a(1-a)$, where $w_0$ and $w_a$ are free parameters, and $E^2(z) = \Omega_{\rm m}(1+z)^{3} +\Omega_{\rm r}(1+z)^{4}+ \Omega_{_\Lambda}(1+z)^{3(1+w_0+w_a)}{\rm exp}(-\frac{3 w_a z}{1+z})$.

There are three interacting dark energy models, i.e., the ${\tilde\Lambda}$CDM model, the e${\tilde\Lambda}$CDM model, and the $\Lambda (t)$CDM model, considered
in this paper. The former two models are motivated from the time-varying gravitational ``constant'' $G$ and the cosmological ``constant'' $\Lambda$, which effectively lead to the interaction between dark energy
and matter. The last one is a phenomenological fluid model with an assumptive direct interaction between dark energy and dark matter, whose interaction term $Q$ is not derived from first principles and its form is purely phenomenological and for the convenience of calculation.


In the next section, we will use the observational datasets to constrain the ${{ \tilde\Lambda}}$CDM, e${{ \tilde\Lambda}}$CDM,
$w$CDM, $\Lambda (t)$CDM, and CPL models from the point of view of alleviating the $H_0$ tension. The results are compared with the base 6-parameter $\Lambda$CDM model that is taken as a benchmark model in this work.

\section{Data and method}\label{sec3}

We summarize the observational data used in this work below.

\begin{table*}
\renewcommand{\arraystretch}{1.5}

\begin{tabular}{|c|c c c c|}

\hline
Model &$\Lambda$CDM&$w$CDM&$\Lambda$(t)CDM& ${\tilde\Lambda}$CDM  \\
\hline
$\Omega_{\rm b}$&$0.0489\pm0.0005$&$0.0480^{+0.0013}_{-0.0012}$&$0.0491^{+0.0010}_{-0.0009}$&$0.0499^{+0.0019}_{-0.0018}$\\
$\Omega_{\rm c}$&$0.2638\pm0.0055$&$0.2606^{+0.0071}_{-0.0067}$&$0.2622^{+0.0094}_{-0.0086}$&$0.2610^{+0.0072}_{-0.0071}$\\
$w$&$-$&$-1.0256^{+0.0364}_{-0.0360}$&$-$&$-$\\
$\beta$&$-$&$-$&$0.0022^{+0.0063}_{-0.0060}$&$-$\\
$\delta_{\rm G}$&$-$&$-$&$-$&$0.0019^{+0.0032}_{-0.0032}$\\
$H_0~[{\rm km~s^{-1}~Mpc^{-1}}]$&$67.70^{+0.44}_{-0.43}$&$68.25^{+0.87}_{-0.89}$&$67.49^{+0.81}_{-0.85}$&$66.95^{+1.39}_{-1.35}$\\
$\Omega_{\rm m}$&$0.3127\pm0.0059$&$0.3087^{+0.0082}_{-0.0077}$&$0.3113^{+0.0097}_{-0.0088}$&$0.3109^{+0.0066}_{-0.0065}$\\
\hline
${H_0}~{\rm tension}$&$4.25\sigma$&$3.46\sigma$&$3.98\sigma$&$3.59\sigma$\\
$\chi_{\rm min}^{2}$&$1043.539   $&$1043.068$&$1042.297$&$1043.201   $\\
$\Delta {\rm AIC}$&$0$&$1.529$&$0.758$&$1.662$\\
$\Delta {\rm  BIC}$&$0$&$6.492$&$5.721$&$6.625$\\
\hline
\end{tabular}

\caption{ The constraint results of parameters in the $\Lambda$CDM model and the one-parameter extension models with the CBS data.}\label{tab1}
\end{table*}

\begin{table*}
\renewcommand{\arraystretch}{1.5}

\begin{tabular}{|c|c c c c|}
\hline
Model &$\Lambda$CDM&$w$CDM&$\Lambda$(t)CDM& ${\tilde\Lambda}$CDM  \\
\hline
$\Omega_{\rm b}$&$0.0483\pm0.0005$&$0.0458^{+0.0011}_{-0.0010}$&$0.0481\pm0.0009$&$0.0452^{+0.0013}_{-0.0012}$\\
$\Omega_{\rm c}$&$0.2569^{+0.0052}_{-0.0050}$&$0.2491^{+0.0058}_{-0.0057}$&$0.2600^{+0.0093}_{-0.0088}$&$0.2688^{+0.0069}_{-0.0072}$\\
$w$&$-$&$-1.0832^{+0.0324}_{-0.0339}$&$-$&$-$\\
$\beta$&$-$&$-$&$-0.0030\pm0.0062$&$-$\\
$\delta_{\rm G}$&$-$&$-$&$-$&$-0.0062^{+0.0025}_{-0.0023}$\\
$H_0~[\rm km~s^{-1}~Mpc^{-1}]$&$68.26\pm0.42$&$69.88^{+0.77}_{-0.76}$&$68.50^{+0.85}_{-0.82}$&$70.69^{+1.06}_{-1.08}$\\
$\Omega_{\rm m}$&$0.3053^{+0.0057}_{-0.0054}$&$0.2949^{+0.0067}_{-0.0066}$&$0.3080^{+0.0094}_{-0.0090}$&$0.3140^{+0.0065}_{-0.0068}$\\
\hline
${H_0}~{\rm tension}$&$3.90\sigma$&$2.57\sigma$&$3.36\sigma$&$1.88\sigma$\\
$\chi_{\rm min}^{2}$&$1061.659   $&$1055.035    $&$1060.435$&$1055.394$\\
$\Delta {\rm AIC}$&$0$&$-4.624$&$0.776$&$-4.265$\\
$\Delta {\rm  BIC}$&$0$&$0.339$&$5.739$&$0.698$\\
\hline
\end{tabular}

\caption{The constraint results of parameters in the $\Lambda$CDM model and the one-parameter extension models with the CBSH data.}\label{tab2}
\end{table*}

\begin{figure*}[!htp]
\includegraphics[scale=0.5]{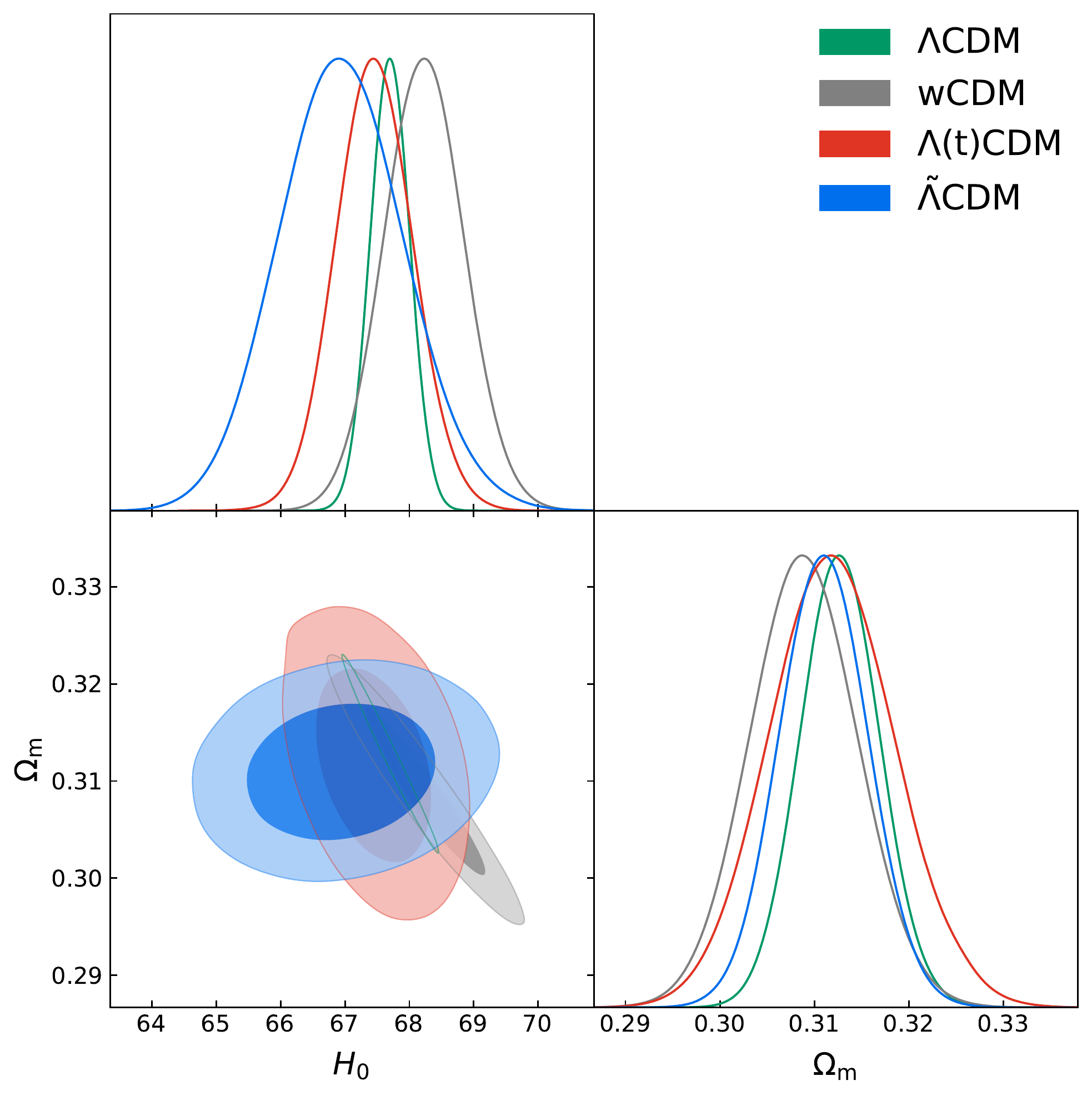}
\centering \caption{\label{fig1} Observational constraints on $H_0$ and $\Omega_{\rm m}$ ($68.3\%$ and $95.4\%$ confidence level) in the $\Lambda$CDM, $w$CDM, $\Lambda$(t)CDM, and ${\tilde\Lambda}$CDM models using the $\rm CBS$ data. Here, $H_0$ is in units of ${\rm km~s^{-1}~Mpc^{-1}}$.}
\end{figure*}

\begin{figure*}[!htp]
\includegraphics[scale=0.5]{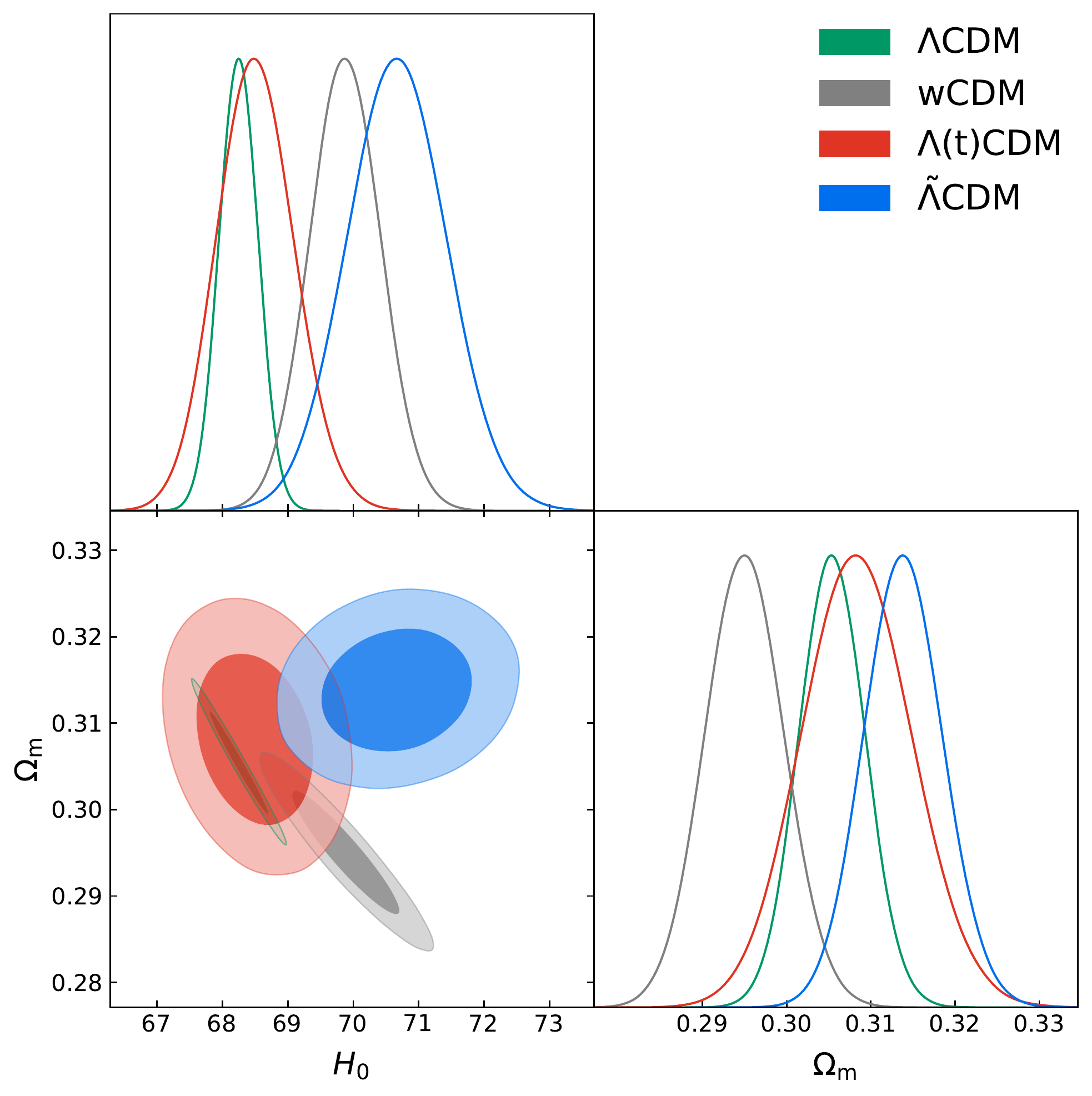}
\centering \caption{\label{fig2} Observational constraints on $H_0$ and $\Omega_{\rm m}$ ($68.3\%$ and $95.4\%$ confidence level) in the $\Lambda$CDM, $w$CDM, $\Lambda$(t)CDM, and ${\tilde\Lambda}$CDM models using the CBSH data combination. Here, $H_0$ is in units of ${\rm km~s^{-1}~Mpc^{-1}}$.}
\end{figure*}

\begin{figure*}[!htp]
\includegraphics[scale=0.55]{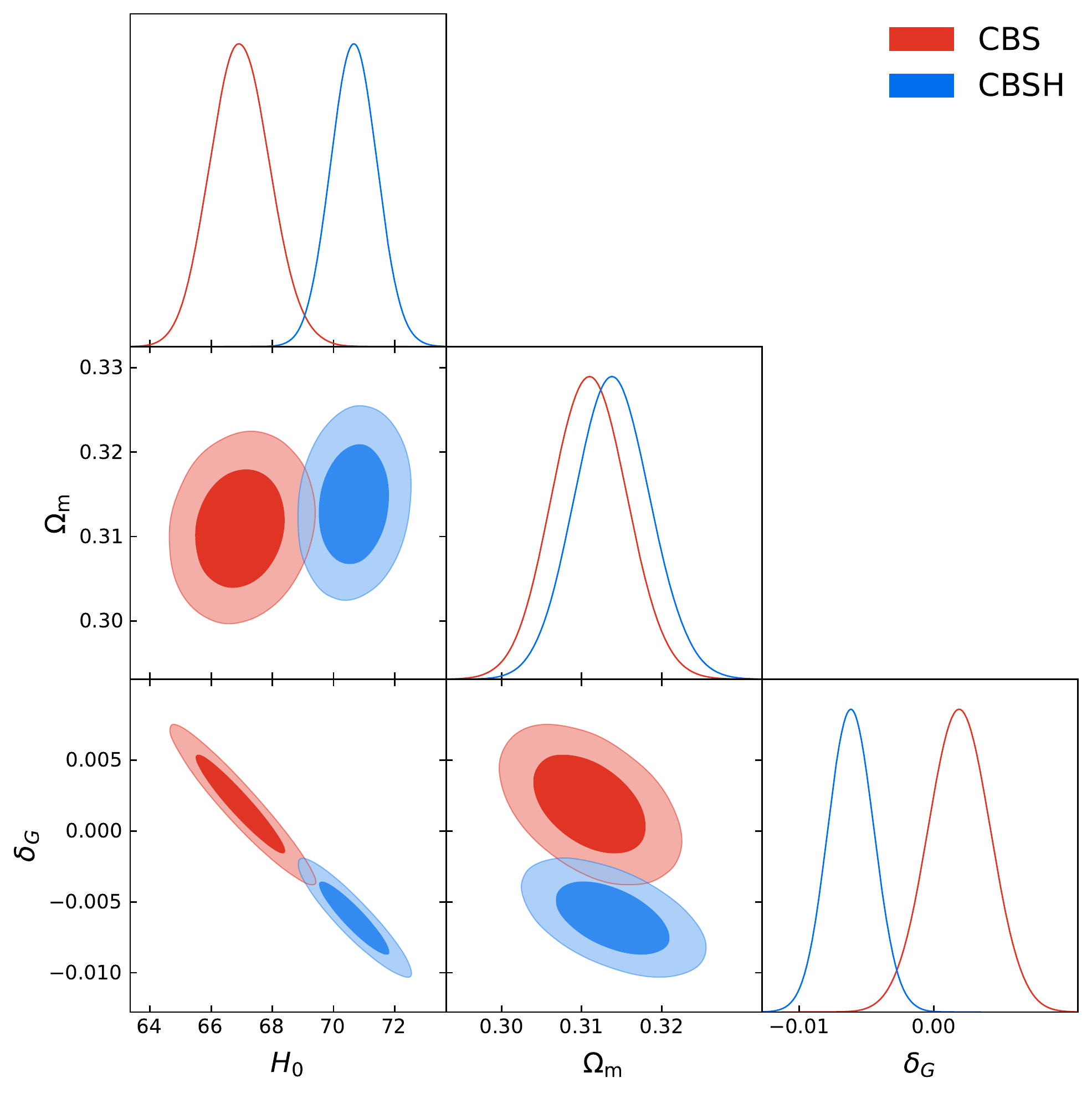}
\centering \caption{\label{fig3} Observational constraints on $H_0$, $\Omega_{\rm m}$, and $\delta_{\rm G}$ ($68.3\%$ and $95.4\%$ confidence level) in the ${\tilde\Lambda}$CDM model using the $\rm CBS$ and $\rm CBSH$ data combinations. Here, $H_0$ is in units of ${\rm km~s^{-1}~Mpc^{-1}}$. }
\end{figure*}

1. CMB: In this work, we use the distance prior data from {\it Planck} 2018 \cite{Chen:2018dbv} for convenience.

2. BAO: The BAO data used in this work include five data points from three observations, i.e., $z_{\rm eff}= 0.016$ from the 6dF Galaxy Survey \cite{Beutler:2011hx}; $z_{\rm eff}= 0.15$ from Main Galaxy Sample of Data Release 7 of Sloan Digital Sky Survey \cite{Ross:2014qpa}; $z_{\rm eff}= 0.38$, $z_{\rm eff}= 0.51$, and $z_{\rm eff}= 0.61$ from the Data Release 12 of Baryon Oscillation Spectroscopic Survey \cite{Alam:2016hwk}.

3. SNIa: We employ the SNIa Pantheon compilation \cite{Scolnic:2017caz} containing 1048 data points.

4. $H_0$: The measurement result of ${H_0} = (74.03 \pm 1.42)~\rm{km~s^{-1} Mpc^{-1}}$ from distance ladder given by the SH0ES team \cite{Riess:2019cxk} is used as a Gaussian prior.

We use the Markov-chain Monte Carlo (MCMC) package {\tt CosmoMC}~\cite{Lewis:2002ah} to perform the cosmological fits. We consider two data combinations in this work, namely, CMB+BAO+SN (abbreviated as CBS) and CBS+$H_0$ (abbreviated as CBSH). It should be emphasized that Bayesian joint analyses cannot automatically show inconsistencies between datasets. However, for the purpose of investigating whether our models can relieve the tension or not, we still combine the local $H_0$ measurement with the CMB+BAO+SN dataset to perform joint analyses as conducted by some other researches \cite{Poulin:2018cxd,Agrawal:2019lmo,Lin:2019qug,DiValentino:2019jae}.


Since the cosmological models have different numbers of free parameters, using only $\chi^2_{\rm min}$ values to compare models is obviously unfair. Thus we use Akaike information criterion (AIC) and Bayesian information criterion (BIC) to perform some punishments to the models having more parameters, which embodies the principle of Occam's Razor to some extent.
We adopt AIC and BIC \cite{Szydlowski:2008by,delCampo:2011jp,Huterer:2016uyq,Liddle:2004nh} given by
\begin{eqnarray}
{\rm AIC}\equiv  \chi^2 +2d \label{aic},\quad {\rm BIC}\equiv  \chi^2 +d\ln N,
\label{aic}
\end{eqnarray}
where $d$ is the number of free parameters and
$N$ is the number of observational data points.
The $\chi^2$ functions for the two data combinations are given by
\begin{eqnarray}
\chi^2&=& \chi^2_{\rm CMB} + \chi^2_{\rm BAO} + \chi^2_{\rm SN},\label{chi}\\
\chi^2 &=& \chi^2_{\rm CMB} + \chi^2_{\rm BAO} + \chi^2_{\rm SN} + \chi^2_{H_0}.
\label{chi0}
\end{eqnarray}

The $\Lambda$CDM model is taken as a benchmark model in the comparison, and thus its AIC and BIC values are set to be zero. For other cosmological models, only the differences from $\Lambda$CDM, $\Delta$AIC$=\Delta\chi^2+2\Delta d$ and $\Delta$BIC$=\Delta\chi^2+\Delta d\ln N$, are important and should be considered.

\begin{figure*}[!htp]
\includegraphics[scale=0.5]{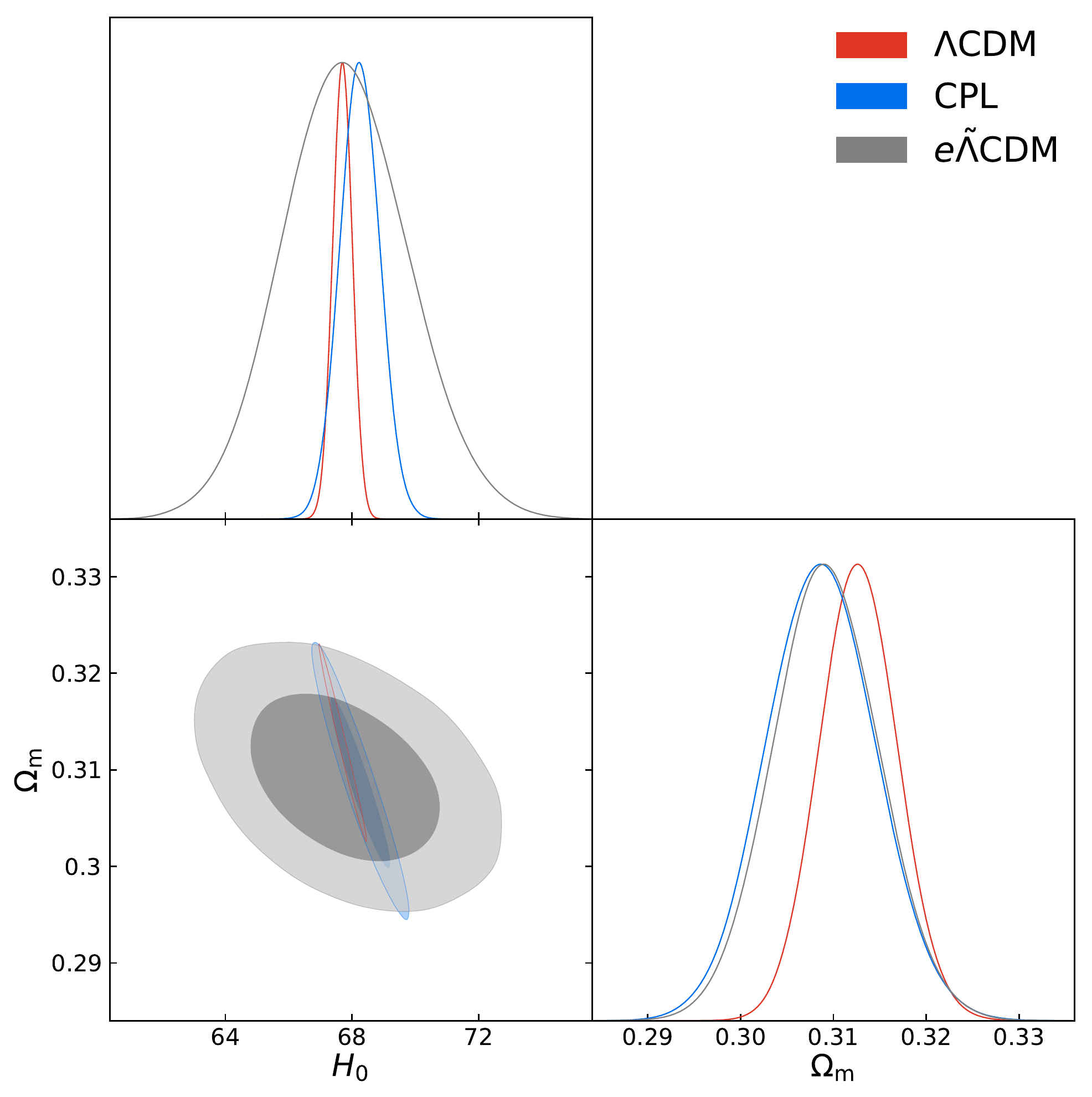}
\centering \caption{\label{fig4} Observational constraints on $H_0$ and $\Omega_{\rm m}$ ($68.3\%$ and $95.4\%$ confidence level) in the $\Lambda$CDM, CPL, and $e{\tilde\Lambda}$CDM models using the $\rm CBS$ data. Here, $H_0$ is in units of ${\rm km~s^{-1}~Mpc^{-1}}$. }
\end{figure*}

\begin{figure*}[!htp]
\includegraphics[scale=0.5]{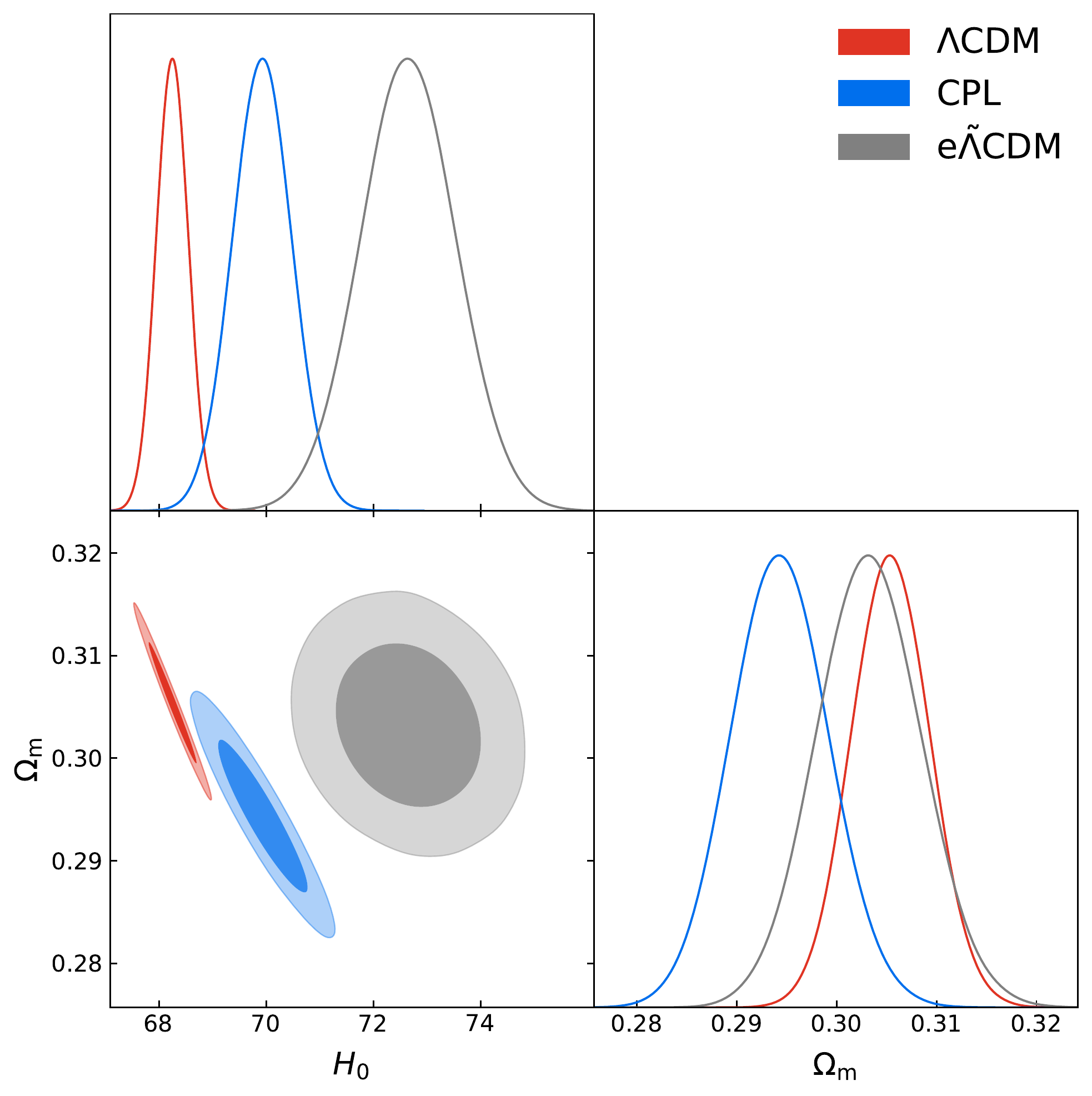}
\centering \caption{\label{fig5} Observational constraints on $H_0$ and $\Omega_{\rm m}$ ($68.3\%$ and $95.4\%$ confidence level) in the $\Lambda$CDM, CPL, and $e{\tilde\Lambda}$CDM models using the $\rm CBSH$ data. Here, $H_0$ is in units of ${\rm km~s^{-1}~Mpc^{-1}}$.}
\end{figure*}

\begin{figure*}[!htp]
\includegraphics[scale=0.65]{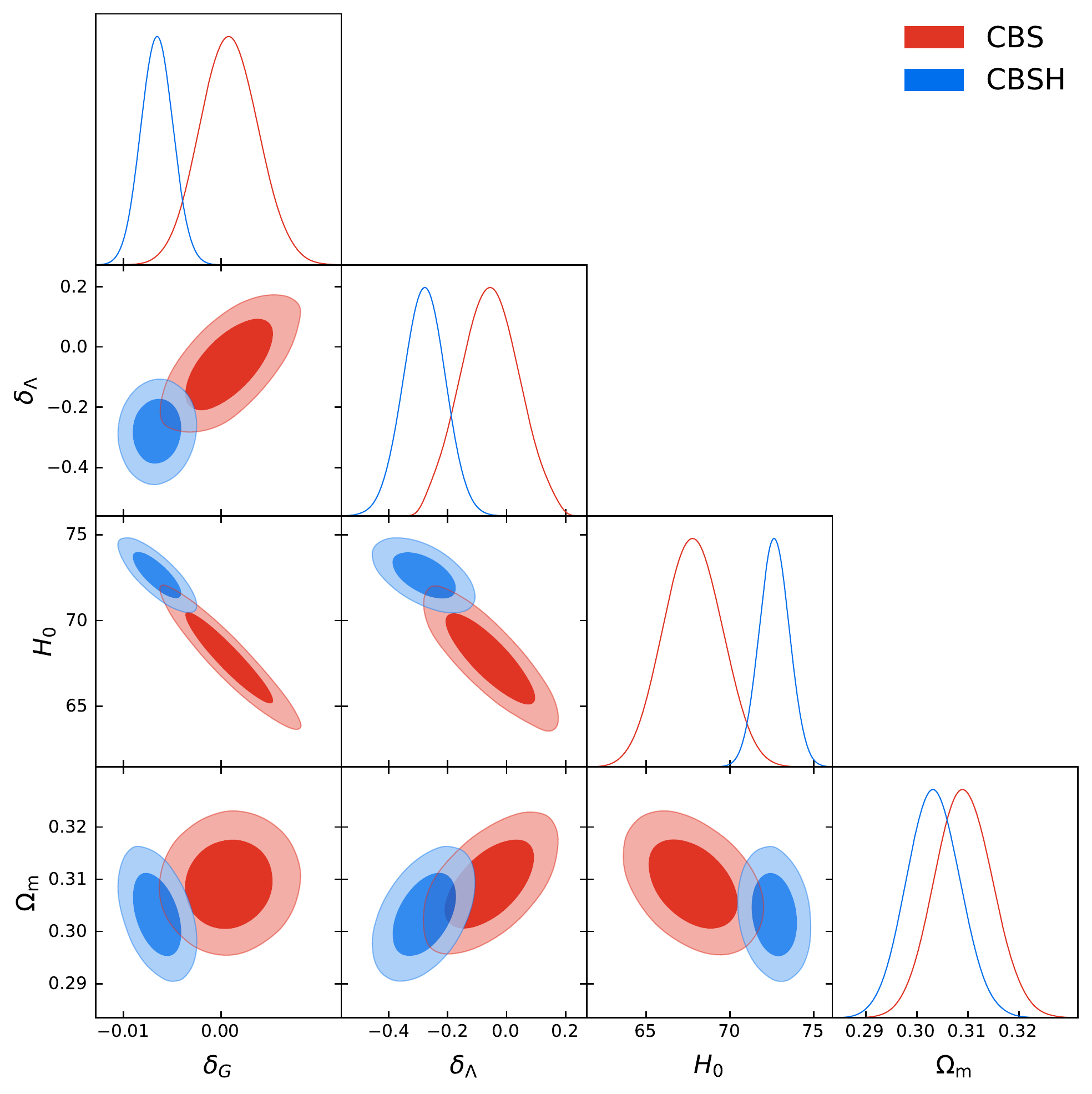}
\caption{Observational constraints ($68.3\%$ and $95.4\%$ confidence level) on $H_0$, $\Omega_{\rm m}$, $\delta_{\rm G}$, and $\delta_\Lambda$ in the $e{\tilde\Lambda}$CDM model using the CBS and CBSH data. Here, $H_0$ is in units of ${\rm km~s^{-1}~Mpc^{-1}}$.}
\label{fig6}
\end{figure*}

\begin{table*}
\renewcommand{\arraystretch}{1.5}

\begin{tabular}{|c|c c|c c|}

\hline
Data &\multicolumn{2}{c|}{$\rm CBS$}&\multicolumn{2}{c|}{${\rm CBSH}$}\\
\hline
Model &$\rm CPL$&${e\tilde\Lambda}$CDM &$\rm CPL$& ${e\tilde\Lambda}$CDM\\

\hline
$\Omega_{b}$&$0.0481^{+0.0012}_{-0.0013}$&$0.0488^{+0.0036}_{-0.0035}$&$0.0457^{+0.0011}_{-0.0010}$&$0.0425^{+0.0015}_{-0.0014}$\\
$\Omega_{c}$&$0.2603^{+0.0073}_{-0.0069}$&$0.2604^{+0.0072}_{-0.0073}$&$0.2478^{+0.0066}_{-0.0054}$&$0.2607^{+0.0072}_{-0.0073}$\\
$w_0$&$-1.0439^{+0.0964}_{-0.0846}$&$-$&$-1.1216^{+0.0930}_{-0.0848}$&$-$\\
$w_a$&$0.0823^{+0.2852}_{-0.3685}$&$-$&$0.1517^{+0.3113}_{-0.3585}$&$-$\\
$\delta_{\rm G}$&$-$&$0.0009^{+0.0042}_{-0.0043}$&$-$&$-0.0066^{+0.0023}_{-0.0022}$\\
$\delta_{\Lambda}$&$-$&$-0.0525^{+0.1365}_{-0.1466}$&$-$&$-0.2832^{+0.1025}_{-0.0966}$\\
\hline
$H_0~[{\rm km~s^{-1}~Mpc^{-1}}]$&$68.23^{+0.90}_{-0.86}$&$67.71^{+2.64}_{-2.40}$&$69.98^{+0.71}_{-0.81}$&$72.69^{+1.23}_{-1.28}$\\
$\Omega_{\rm m}$&$0.3084^{+0.0083}_{-0.0080}$&$0.3092^{+0.0078}_{-0.0081}$&$0.2935^{+0.0075}_{-0.0062}$&$0.3031^{+0.0073}_{-0.0073}$\\
\hline
${H_0}~{\rm tension}$&$3.47\sigma$&$2.18\sigma$&$2.51\sigma$&$0.71\sigma$\\
$\chi_{\rm min}^{2}$&$1043.045  $&$1043.037  $&$1054.865   $&$1047.409    $\\
$\Delta {\rm AIC}$&$3.498$&$3.501$&$-2.794$&$-10.250$\\
$\Delta {\rm  BIC}$&$13.431$&$13.423$&$7.133$&$-0.323$\\

\hline
\end{tabular}

\caption{\label{tab3} The constraint results of parameters in the two-parameter extension models with the CBS and CBSH data.}
\end{table*}

\section{Results and discussion}\label{sec4}

We show the posterior distributions of cosmological parameters in the $\Lambda$CDM model and the one-parameter extensions to $\Lambda$CDM in Figs. \ref{fig1}--\ref{fig3} and report the detailed results in Tabs.~\ref{tab1} and \ref{tab2}.

Fig.~\ref{fig1} and Table~\ref{tab1} show the results of using the CBS data to constrain the $\Lambda$CDM model and its one-parameter extensions, i.e., $w$CDM, $\Lambda(t)$CDM, and ${\tilde\Lambda}$CDM. We can see that, in this case, only $w$CDM can slightly alleviate the $H_0$ tension, with the best-fit value of $H_0$ equal to 68.25 km s$^{-1}$ Mpc$^{-1}$; $\Lambda(t)$CDM and ${\tilde\Lambda}$CDM even get smaller $H_0$ values (best-fit values), and they are equal to $67.49^{+0.81}_{-0.85}$ km s$^{-1}$ Mpc$^{-1}$ and $66.95^{+1.39}_{-1.35}$ km s$^{-1}$ Mpc$^{-1}$, respectively. This is because using the CBS data leads to the results (central values) of $w<-1$ in $w$CDM, $\beta>0$ in $\Lambda(t)$CDM, and $\delta_G>0$ in ${\tilde\Lambda}$CDM. It is known that the phantom energy case of $w<-1$ can lead to a larger $H_0$. The cases of $\beta>0$ in $\Lambda(t)$CDM and $\delta_G>0$ in ${\tilde\Lambda}$CDM do not realize an effective phantom, but on the contrary they actually realize an effective quintessence, and thus in this situation $\Lambda(t)$CDM and ${\tilde\Lambda}$CDM cannot effectively alleviate the $H_0$ tension. We can see from Fig.~\ref{fig1} that basically both $\Lambda(t)$CDM and ${\tilde\Lambda}$CDM are in good agreement with the $\Lambda$CDM cosmology in the case of CBS constraint. In addition, from Table~\ref{tab1} we find that $\Lambda$CDM is the best one in fitting to the CBS data, and the other three extension models actually cannot provide a good fit to the CBS data, which can be seen from their large values of $\Delta {\rm AIC}$ and $\Delta {\rm BIC}$.

However, when the $H_0$ direct measurement from the SH0ES team is added in the data combination, the situation will be dramatically changed. Now, we consider the ${\rm CBS}+H_0$ data combination (also abbreviated as CBSH), and the constraint results are shown in Fig.~\ref{fig2} and Table~\ref{tab2}. We find that in this case $w$CDM and ${\tilde\Lambda}$CDM can yield larger values of $H_0$, but $\Lambda(t)$CDM still cannot make $H_0$ larger. Actually, even though the $H_0$ prior is involved in the data combination, one cannot detect the coupling between vacuum energy and cold dark matter in $\Lambda(t)$CDM; the constraint on $\beta$ is $\beta=-0.0030\pm 0.0062$. Therefore, $\Lambda(t)$CDM cannot help alleviate the $H_0$ tension (the tension is still in 3.36$\sigma$). Although $w$CDM slightly prefers a phantom energy with $w=-1.0832^{+0.0324}_{-0.0339}$, and the anti-correlation between $w$ and $H_0$ can help relieve the $H_0$ tension, it still cannot lead to a large enough value of $H_0$; it gives $H_0=69.88^{+0.77}_{-0.76}$ km s$^{-1}$ Mpc$^{-1}$, and the tension is still in 2.57$\sigma$. Evidently, the focus is on ${\tilde\Lambda}$CDM. When the $H_0$ prior is added in the data combination, ${\tilde\Lambda}$CDM yields a much larger $H_0$, i.e., $H_0=70.69^{+1.06}_{-1.08}$ Mpc$^{-1}$, leading to the $H_0$ tension enormously relieved (the tension is now in 1.88$\sigma$). This is owing to the fact that a non-zero $\delta_G$ is obtained in this case, i.e., $\delta_{\rm G} =-0.0062^{+0.0025}_{-0.0023}$. A negative $\delta_G$ implies that the ``cosmological constant'' in ${\tilde\Lambda}$CDM becomes larger and larger with the cosmological evolution, and thus actually it is an effective phantom providing stronger repulsive force driving the cosmic acceleration. The faster late-time cosmic expansion means a larger $H_0$, and thus a more negative $\delta_G$ will yield a larger $H_0$.

In Fig.~\ref{fig3}, we compare the constraints from CBS and CBSH on ${\tilde\Lambda}$CDM. We can clearly see that, when the $H_0$ prior is added, the situation is dramatically changed, as the value of $\delta_G$ is changed from the case of consistent with 0 to the one with a negative value. The anti-correlation between $\delta_G$ and $H_0$ is also explicitly shown, and we can immediately find that a negative $\delta_G$ leads to a high value of $H_0$. In the cases of CBS and CBSH, the $H_0$ tension is in 3.59$\sigma$ and 1.88$\sigma$, respectively. In addition, it is easily found that, in the CBS case, ${\tilde\Lambda}$CDM is not favored, but in the CBSH case, ${\tilde\Lambda}$CDM is strongly preferred (see the negative values of $\Delta$AIC and $\Delta$BIC in Table~\ref{tab2}). Therefore, for ${\tilde\Lambda}$CDM, we find that $\delta_G$ is very sensitive to $H_0$, and the local measurement of $H_0$ in the datasets becomes a dominant factor in the cosmological fit. But for $\Lambda(t)$CDM, the coupling parameter $\beta$ is not sensitive to $H_0$. We can thus conclude that ${\tilde\Lambda}$CDM as a kind of interacting dark energy model behaves much better than $\Lambda(t)$CDM in the sense of resolving the $H_0$ tension.

Next, let us see the situation of the two-parameter extension models, i.e., the CPL and e${\tilde\Lambda}$CDM models. The main results are shown in Figs.~\ref{fig4}--\ref{fig6} and Table~\ref{tab3}. The comparison of CPL and e${\tilde\Lambda}$CDM is given in Figs.~\ref{fig4} and \ref{fig5}; Fig.~\ref{fig4} shows the case of CBS and Fig.~\ref{fig5} shows the case of CBSH. From Fig.~\ref{fig4}, we find that in the CBS case neither CPL nor e${\tilde\Lambda}$CDM can effectively alleviate the $H_0$ tension. In this case in e${\tilde\Lambda}$CDM both $\delta_G$ and $\delta_\Lambda$ are well consistent with 0, and thus the value of $H_0$ cannot be increased (see Table~\ref{tab3} for detailed results). From Fig.~\ref{fig5}, we find that, once the $H_0$ prior is added in the data combination, the situation for e${\tilde\Lambda}$CDM is changed dramatically. In this case, we have $\delta_{G}=-0.0066^{+0.0023}_{-0.0022}$ and $\delta_{\Lambda}=-0.2832^{+0.1025}_{-0.0966}$, showing the results of $\delta_G<0$ and $\delta_\Lambda<0$ at the more than 2$\sigma$ level. Hence, e${\tilde\Lambda}$CDM in the CBSH case can also yield an effective phantom behavior, which leads to a high value of $H_0$, i.e., $H_0=72.69^{+1.23}_{-1.28}$ km s$^{-1}$ Mpc$^{-1}$. Therefore, in the CBSH case, e${\tilde\Lambda}$CDM can well resolve the $H_0$ tension, with the tension relieved to 0.71$\sigma$ level. The comparison of the values of $\Delta {\rm AIC}$ and $\Delta {\rm BIC}$ is explicitly shown in Table~\ref{tab3}, and we can see that the e${\tilde\Lambda}$CDM model in the CBSH case is the best one (with $\Delta {\rm AIC}= -10.250$ and $\Delta {\rm BIC}= -0.323$) in the sense of both relieving the $H_0$ tension and fitting to the observational data. In Fig.~\ref{fig6}, for the constraints on e${\tilde\Lambda}$CDM, we make a comparison for the cases of CBS and CBSH. From the posterior distributions of $\delta_G$, $\delta_\Lambda$, and $H_0$, we can clearly see their shifts after the addition of the $H_0$ prior into the data combination.


\section{Robustness of results}\label{sec5}

There may still be some concerns about the robustness of our results. The first concern could arise from the belief that baryons and radiation should receive less modifications than cold dark matter, i.e., the interaction between dark energy and radiation (or baryons) is tightly constrained. Using the CBS and CBSH datasets, we
make several attempts to study how different values of $\delta_G$ associate to the matter
and radiation terms in $E^2(z)$ in Eq.~(\ref{final}), to illustrate the effects of different components on the results. Indeed, we find that the coupling of cold dark matter and dark energy plays an important role in the interaction between matter and dark energy. Therefore, in this article, we present the results of the particular case in which only cold dark matter and dark energy are interacting,
\begin{equation}
E^2(z)
= \Omega_{\rm c}(1+z)^{(3-\delta_{\rm G})} +\Omega_{\rm b}(1+z)^{3} +\Omega_{\rm r}(1+z)^{4}+\Omega_{_\Lambda}(1+z)^{\delta_\Lambda}.
\label{final2}
\end{equation}
{Namely, we only consider the corrections on the evolutions of cold dark matter and dark energy, and assume that $\delta_{\rm G}$ and $\delta_\Lambda$ are independent of each other, so the resulting model can be considered as a limiting case of the e${\tilde\Lambda}$CDM model. Hereafter, this limiting e${\tilde\Lambda}$CDM model is abbreviated as l${\tilde\Lambda}$CDM.}

\begin{table*}
\begin{center}
\renewcommand{\arraystretch}{1.5}
\begin{tabular}{|c|c c|}
\hline
Data & $\rm CBS$&$\rm CBSH$ \\
\hline
$\delta_{\rm G}$&$ 0.0007^{+0.0050}_{-0.0047}$&$-0.0071^{+0.0041}_{-0.0037}$\\
$\delta_{\Lambda}$&$ -0.0595^{+0.1431}_{-0.1403}$&$-0.3442^{+0.1143}_{-0.1056}$\\
$H_0~[{\rm km~s^{-1}~Mpc^{-1}}]$&$ 68.06\pm 1.36$&$71.10^{+0.94}_{-1.07}$\\
$\Omega_{\rm m}$&$  0.3090\pm0.0080$&$ 0.2973^{+0.0073}_{-0.0063}$\\
\hline
\end{tabular}
\caption{\label{tab4}  The constraint results of $\delta_{\rm G}$, $\delta_{\Lambda}$, $H_0$, and $\Omega_{\rm m}$ in the l${\tilde\Lambda}$CDM model with the CBS and CBSH datasets.}
\end{center}
\end{table*}

The constraint results using the CBS and CBSH datasets are listed in Table~\ref{tab4}. The l${\tilde\Lambda}$CDM model gives $H_0= (68.06\pm 1.36)$ km s$^{-1}$ Mpc$^{-1}$ with the CBS dataset and a relatively larger value $H_0= 71.10^{+0.94}_{-1.07}$ km s$^{-1}$ Mpc$^{-1}$ with the CBSH dataset. The $H_0$ tension is relieved to $1.68\sigma$ with the CBSH dataset. This result implies that the interaction of dark energy and cold dark matter plays a dominant role in the background evolution of the e${\tilde\Lambda}$CDM model. Therefore, our models can still be effective in resolving the $H_0$ tension, even if the modifications of the evolutions of radiation and baryons are negligible.

The second concern is that we used the CMB distance prior to constrain the models rather than the full power spectrum of {\it Planck} 2018. In the following, we test the difference of these two data in constraining the e${\tilde\Lambda}$CDM model.
We use the MontePython code \cite{Audren:2012wb} to perform the MCMC analysis.
We also use the two data combinations as above, i.e., CBS (full spectrum) and CBSH (full spectrum), in which the {\it Planck} TT,TE,EE+lowE+lensing data \cite{Aghanim:2018eyx} are used as the CMB data, and other cosmological data are still the same as in Section \ref{sec3}.

We list the results in Table~\ref{tab5} and compare them with the previous results using distance prior of CMB in Table~\ref{tab3}. We find that the mean values of parameters slightly shift and the errors greatly shrink. For the e${\tilde\Lambda}$CDM model, the CBS and CBS (full spectrum) datasets can give $H_0=67.71^{+2.64}_{-2.40}$ km s$^{-1}$ Mpc$^{-1}$ and $H_0=(68.17\pm 0.87)$ km s$^{-1}$ Mpc$^{-1}$, respectively; the CBSH and CBSH (full spectrum) datasets can give $H_0=72.69^{+1.23}_{-1.28}$ km s$^{-1}$ Mpc$^{-1}$ and $H_0=(73.05\pm 0.56)$ km s$^{-1}$ Mpc$^{-1}$, respectively. These results show that although the full power spectrum data of CMB can provide more information than the distance prior, the main conclusions still hold.

There is another tension between the {\it Planck} base-$\Lambda$CDM cosmology and galaxy clustering of the matter fluctuations.
As a result of the full CMB anisotropies data, the amplitude of the matter power spectrum $\sigma_8$ and its related parameter $S_8=\sigma_8(\Omega_m/0.3)^{0.5}$ can be constrained and the $\sigma_8$/$S_8$ tension can also be evaluated.

We discuss the $\sigma_8$/$S_8$ tension in the CBSH (full spectrum) dataset, because the e${\tilde\Lambda}$CDM model can effectively relieve the $H_0$ tension in this dataset. The CBSH (full spectrum) dataset gives $\sigma_8= 0.8099\pm 0.0060$ and $S_8 =0.8194\pm 0.0101$ in the $\Lambda$CDM model and $\sigma_8=  0.8720\pm 0.0090$ and $S_8 =0.8310\pm 0.0110$ in the e${\tilde\Lambda}$CDM model. We find that the value of $\sigma_8$ in the e${\tilde\Lambda}$CDM model increases than that in the $\Lambda$CDM model, but the value of $S_8$ only slightly changes because $\Omega_{\rm m}$ tends to decrease in the e${\tilde\Lambda}$CDM model. We compare with the results from the combination of the KiDS/Viking and SDSS data, $\sigma_8= 0.760^{+0.025}_{-0.020}$ and $S_8= 0.766^{+0.020}_{-0.014}$ \cite{Heymans:2020gsg}.
{In the $\Lambda$CDM model, the $\sigma_8$ and $S_8$ tensions are in $2.36\sigma$ and $2.44\sigma$, respectively, while the ones in the e${\tilde\Lambda}$CDM model are in $4.26\sigma$ and $3.21\sigma$, respectively.}

As a crosscheck, we also compare our constraint results with the results in the literature \cite{Aghanim:2018eyx,Guo:2018ans,Camarena:2021jlr} and find that they are statistically consistent.
{For example, in the $w$CDM model, we obtain $H_0= 68.25^{+0.87}_{-0.89}$ km s$^{-1}$ Mpc$^{-1}$ using the CBS data as shown in Table \ref{tab1}, and Ref. \cite{Aghanim:2018eyx} gives $H_0=(68.34\pm 0.81)$ km s$^{-1}$ Mpc$^{-1}$ using the {\it Planck} 2018 TT,TE,EE+lowE+lensing+BAO+Pantheon data.}
{Moreover, in the $\Lambda$(t)CDM model, we obtain $H_0= 68.50^{+0.85}_{-0.82}~\rm km~s^{-1}~Mpc^{-1}$ using the CBSH data as shown in Table \ref{tab2}, and Ref. \cite{Guo:2018ans} gives $H_0= (69.36 \pm 0.82)~\rm km~s^{-1}~Mpc^{-1}$ using also the CBSH data, but in which the {\it Planck} 2015 data and an earlier local $H_0$ measurement are used.}
Through all these tests of the robustness of the results, we further confirm that our models are helpful to relieve the $H_0$ tension.

\begin{table*}
\renewcommand{\arraystretch}{1.5}

\begin{tabular}{|c|c c|c c|}

\hline
Data &\multicolumn{2}{c|}{CBS (full spectrum)}&\multicolumn{2}{c|}{CBSH (full spectrum)}\\

\hline
Model & $\Lambda$CDM &$e{\tilde\Lambda}$CDM model &  $\Lambda$CDM &$e{\tilde\Lambda}$CDM model  \\

\hline
$\delta_{\rm G}$&$-$&$ 0.00030\pm 0.00101$&$-$&$-0.00387^{+0.00054}_{-0.00072}$\\
$\delta_{\Lambda}$&$-$&$-0.1102\pm 0.1201$&$-$&$ -0.2511^{+0.0162}_{-0.0183}$\\
$H_0~[{\rm km~s^{-1}~Mpc^{-1}}]$&$67.71\pm 0.41$&$68.17\pm 0.87$&$68.01\pm 0.40$&$73.05\pm 0.56$\\
$\Omega_{\rm m}$&$0.3108\pm 0.0055$&$ 0.3075\pm 0.0080$&$0.3068\pm 0.0053$&$ 0.2724^{+0.0052}_{-0.0045}$\\
$\sigma_{8}$&$0.8111\pm 0.0061$&$0.8150\pm 0.0121$&$0.8099\pm 0.0060$&$ 0.8720\pm 0.0090$\\
$S_{8}$&$0.8263\pm 0.0102$&$0.8252\pm 0.0132$&$0.8194\pm 0.0101$&$0.8310\pm 0.0110$\\
\hline
\end{tabular}

\caption{\label{tab5}  The constraint results of $\delta_{\rm G}$, $\delta_{\Lambda}$, $H_0$, $\Omega_{\rm m}$, $\sigma_{8}$, and $S_{8}$ in the $\Lambda$CDM model and the $e{\tilde\Lambda}$CDM model with the CBS (full spectrum) and CBSH (full spectrum) datasets.}
\end{table*}

\section{Conclusion}\label{sec6}

In this work, we consider a phenomenological cosmological model motivated by the asymptotic safety of gravitational field theory. In this model, the matter and radiation densities and the cosmological constant receive a correction parametrized by the parameters $\delta_G$ and $\delta_\Lambda$, leading to that both the evolutions of the matter and radiation densities and the cosmological constant slightly deviate from the standard forms. Actually, this model can be explained by the scenario of vacuum energy interacting with matter and radiation. Furthermore, we consider two cases of the model: (i) ${\tilde\Lambda}$CDM with one additional free parameter $\delta_G$, in which $\delta_{\rm G}$ and $\delta_\Lambda$ are related by a low-redshift limit relation and (ii) e${\tilde\Lambda}$CDM with two additional free parameters $\delta_G$ and $\delta_\Lambda$ independent of each other. We use the current observational data (CBS and CBSH) to constrain the models.

We find that, when using the CBS data, neither ${\tilde\Lambda}$CDM nor e${\tilde\Lambda}$CDM can effectively alleviate the $H_0$ tension. In this case, we obtain that both $\delta_G$ and $\delta_\Lambda$ are around 0, and thus the models are well consistent with $\Lambda$CDM. Actually, in this case, the CBS data prefer $\Lambda$CDM more over ${\tilde\Lambda}$CDM and e${\tilde\Lambda}$CDM.

However, when the direct measurement of $H_0$ by the SH0ES team is added in the data combination (i.e., CBSH is considered), the situation is dramatically changed. We find that in this case both $\delta_G<0$ and $\delta_\Lambda<0$ are obtained at the more than 2$\sigma$ significance.
We find that, when using the CBSH data to constrain ${\tilde\Lambda}$CDM and e${\tilde\Lambda}$CDM, the $H_0$ tension can be greatly relieved. In particular, for example, in the case of e${\tilde\Lambda}$CDM, the $H_0$ tension can be resolved to 0.71$\sigma$. In addition, through an analysis of model selection using the information criteria, we find that the CBSH data prefer e${\tilde\Lambda}$CDM over $\Lambda$CDM. We also perform some tests on the robustness of our results, including a limiting case in which the modifications of the evolutions of radiation and baryons are negligible, a comparison of using the CMB distance prior and the full power spectrum of {\it Planck} 2018 to constrain parameters, and a crosscheck with the previous results in other works. These tests confirm our results, so we can conclude that, from a comprehensive analysis, e${\tilde\Lambda}$CDM as an interacting dark energy model is much better than $\Lambda(t)$CDM in the sense of both relieving the $H_0$ tension and fitting to the current observational data.

\begin{acknowledgments}

This work was supported by the National Natural Science Foundation of China (Grant Nos. 11975072, 11875102, 11835009, and 11690021), the Liaoning Revitalization Talents Program (Grant No. XLYC1905011), the Fundamental Research Funds for the Central Universities (Grant No. N2005030), and the Top-Notch Young Talents Program of China (Grant No. W02070050).

\end{acknowledgments}

\end{document}